# Impacts of Fermi Level Pinning at Hole-Selective Contacts in CdSeTe/CdTe Solar Cells

Ariful Islam[1], Nathan D. Rock[2], Kh. Aaditta Arnab[2], Nicholas Miller[3], James Becker[3], and Michael A. Scarpulla[1,2]

[1.] Electrical and Computer Engineering, University of Utah, Salt Lake City, USA

[2.] Materials Science and Engineering, University of Utah, Salt Lake City, USA

[3] First Solar, Inc., Perrysburg, OH, USA

*Corresponding author:* Michael A. Scarpulla (mike.scarpulla@utah.edu)

**Abstract:** P-type doped CdTe free surfaces, Schottky contacts, and even interfaces with isostructural p-ZnTe frequently exhibit downward band bending and moderate to high recombination velocities. Fermi level pinning by donor-like states can explain these band diagram features, as well as device response characteristics such as 1$^{st}$ quadrant rollover in current-voltage (JV) versus temperature (JVT). Parasitic downward band bending also produces voltage-dependent photocurrent collection, producing fill factor (FF) efficiency losses, JV dark/light non-superposition (or "JV take-off"), and irregularities in $J_{sc}$-$V_{oc}$ and Suns-$V_{oc}$ measurements. Herein, we develop a device physics model of state-of-the-art CdSeTe/CdTe solar cells consistent with known characterization of materials and devices, including the optical, thermalization, and trapping effects of band tail states and isolated defects. We use this model to demonstrate that Fermi-level pinning at the p-ZnTe/p-CdSeTe hole contact by donor-like defects reproduces the aforementioned observables, and conclude that (for contemporary few-µm absorber thicknesses and low mobilities) it primarily affects FF rather than $V_{oc}$. We investigate the performance gains possible from hypothetical passivated, hole-selective layers at the ZnTe/CdTe interface, which eliminate the downwards band bending caused by donor-like defects. For thinner devices and larger minority carrier diffusion lengths, these strategies will become more important for continued efficiency improvements.

**Keywords:** CdSeTe/CdTe solar cells, interface recombination, dark-light non-superposition, JV lift-off, JV take-off, Fermi level pinning, hole selective contact.

**Introduction:**

Cadmium telluride (CdTe) with a bandgap of 1.5 eV is the highest-volume commercial thin-film polycrystalline photovoltaic technology with over 25 $GW_{DC}$ installed, accounting for approximately 34% of US utility-scale installations and 3% of global installed capacity[1], [2], [3]. Recent advances have eliminated CdS buffer layers and incorporated selenium (Se) to form graded $CdSe_{1-x}Te_x$ (CdSeTe) absorbers, reducing the bandgap to ~1.40 eV and enhancing short circuit current ($J_{sc}$)[2], [4]. While short circuit current ($J_{sc}$) in CdTe-based solar cells has approached theoretical limits for AM1.5 insolation, the open-circuit voltage ($V_{oc}$) remains significantly below the radiative limit, as does the fill factor (FF), reducing the overall device efficiency [5].

A known challenge for solar cells with CdTe-based absorbers, exacerbated by difficulties in p-type dopability, is the formation of p-type Ohmic, hole-selective contacts. Free surfaces, grain boundaries, and interfaces to metals, as well as isostructural p-ZnTe, tend to form downward band bending with an approximate barrier magnitude of 0.1-0.4 eV, hinting at a possible universal origin of Fermi level pinning from donor-like states. Pinning for n-type doped CdTe would suggest acceptor-like states, and it is common to represent complicated energy distributions of interface states using amphoteric defects with densities and transition level(s) (although this is not the only way to pin $E_F$) [6]. Other observations indicate that electronegative Cl (which tends to pair with ionized, positively charged donor-like defects) effectively passivates such surfaces and interfaces. Additionally, Se, which is more-electronegative than Te exhibits passivating effects for point defects and grain boundaries as evidenced by longer minority carrier lifetimes in CdTe-based x-ray detectors and thin film solar cells [7], [8], [9]. The tendency of Te to dimerize and form donor-like states in the bandgap when undercoordinated by Cd could provide a universal root cause for these phenomena. The explanation, common in the literature, that difficulties with Ohmic contact formation to p-CdTe are caused by metal work functions and the CdTe valence band being energetically deep, cannot *alone* be the root cause. Besides the known aforementioned Fermi level pinning at CdTe grain boundaries, causing barriers at the grain boundaries [10], GaAs has a similar bandgap (1.42 eV) and electron affinity (4.1 eV, compared to 4.3 for CdTe), yet Ohmic p-type contacts are possible for GaAs [11].

Wider-bandgap, p$^+$-doped ZnTe (~2.3 eV) has proven to be an effective hole-selective contact that blocks electrons by virtue of its large Type I 0.8-0.95 eV conduction band offset ($\Delta E_c$) and favorable valence band alignment $\Delta E_v$ = 0 to -0.15 eV [12], [13]. However, abrupt CdTe/ZnTe interfaces are usually unpassivated (recombination velocity S~$10^2$-$10^4$ cm/s depending on chlorine heat treatment CHT), probably related to dislocations required to accommodate the lattice mismatch of 5.9% [14], [15], [16]. Commonly used sputter deposition of ZnTe may also introduce some damage, although alternative ZnTe deposition methods, such as thermal evaporation, have been anecdotally reported not to eliminate the interface RV.

In this work, we found that certain CdSeTe-based cells produced in First Solar internal research (and not necessarily representing commercially available products) exhibited dark-light JV non-superposition[17], [18], [19], [20], [21], [22], [23]. JV non-superposition occurs when the illuminated J(V) curve, shifted by $J_{sc}$, does not overlap the dark J(V) curve. It is seen in some subsets of fabricated cells but not always; however, it proved to be a decisive clue, allowing us to infer the presence of defects at the p$^+$-ZnTe/p-CdTe interface. In further discussions herein, we adopt $J_{light}$(V) and $J_{dark}$(V) to indicate J(V) measured in light and dark (respectively) and define the collected photocurrent $J_{pc}$(V) as the difference $J_{pc}$(V) = $J_{dark}$(V) - $J_{light}$(V) to allow for the possibility of it being voltage-dependent. The shape of J(V) curves are described within a non-ideal, single-diode equivalent circuit model by 2$^{nd}$ level parameters affecting FF like series resistance ($R_s$), shunt resistance ($R_{sh}$), dark leakage current $J_{o,dark}$, and ideality factor A. Solar cells exhibiting superposition would have $J_{pc}$(V) = constant = $J_{sc}$, which implies that the shape of the J(V) curve is not distorted between dark and light and thus $R_s$, $R_{sh}$, $J_o$, and A all have the same voltage dependencies (or lack thereof) in the dark and light.

Non-superposition can have many origins, but to first order can be coarsely grouped by a) light-induced changes in the device's radiative and non-radiative recombination currents captured by $J_{o,dark}$ and non-ideality factor A, or b) light-induced changes in $R_s$, $R_{sh}$, A, or $J_{pc}$(V). Differences in dark- and light-trap occupation can lead to variations in non-radiative recombination, which, in turn, affect the $J_o$(V) and A factors. For example, traps may become saturated at high illumination intensity and act less effectively as recombination centers. Internal and contact-related transport inefficiencies would cause effects under category b). Changes in either $J_{dark}$(V) or $J_{pc}$(V) could alter $V_{oc}$ = $k_B T$ ln ($J_{pc}/J_{dark}$); however, we demonstrate herein for modern ZnTe/CdTe/CdSeTe devices that even a small hole extraction barrier at the ZnTe/CdTe interface can cause dark-light JV non-superposition and FF losses. We show explicitly that the non-superposition can be caused by $E_F$ pinning at the ZnTe/CdTe interface, and that this does not materially affect recombination at the back contact and thus $V_{oc}$: this observation eliminates possibility a).

We demonstrate that the downwards band bending and dark-light JV non-superposition is driven by donor-like defects at the back p-ZnTe/p-CdTe interface. Contrary to the intuition that recombination would be higher because minority electrons can reach the interface defects more easily, the accompanying downwards band bending suppresses recombination by suppressing hole density at those defects. Simultaneously, the combination of low p-type doping in ZnTe and CdSeTe, low mobility in CdSeTe, and the defect-induced downwards band bending produces lower and voltage-dependent hole collection, which can perhaps be most easily visualized as a voltage-dependent $R_s$ in terms of 2$^{nd}$ level metrics [24]. Neither acceptor-like defects at the back interface nor front-interface defects can reproduce JV non-superposition and 1$^{st}$ quadrant JVT roll-over [25]. However, removing the ZnTe and imposing equivalent downward band bending via a Schottky boundary condition likewise produces J–V non-superposition. Thus, the effect depends primarily on the presence and magnitude of the hole barrier, which is caused by Fermi-level pinning at the back of the absorber, rather than recombination activity at the interface.

**Methods**

To investigate the impact of the back-interface hole barrier and series resistance on CdSeTe/CdTe solar-cell performance, fully processed CdSeTe/CdTe mini-modules fabricated by First Solar were examined. The modules consisted of strings of eight monolithically integrated cells connected in series, representing typical commercial module architecture. Modern CdTe modules are fabricated by monolithic series interconnection on a glass superstrate, employing laser patterning (P1–P2–P3) to define individual cells and achieve high-throughput production [26]. Current–voltage (J–V) characteristics were measured under calibrated AM 1.5G illumination (100 mW cm$^{-2}$) using an in-house class AAA solar simulator. To extract the single-cell performance, the total open-circuit voltage ($V_{oc}$, module) was divided by the number of series-connected cells (N = 8), yielding the $V_{oc}$ value per cell. The short-circuit current ($I_{sc}$, module) was normalized by the measured active area of one cell, giving $J_{sc}$. Before the measurement, a wake-up procedure was carried out by holding the devices at 80° C and 1 sun illumination (100 mW/cm²) for 6 hours (the minimum time required was not investigated herein). After the procedure, $J_{sc}$ and other JV characteristics were significantly higher; for example, before wake-up, the $J_{sc}$ was reduced by nearly 50%. This seems to be an unusual manifestation of sleep/wake-up behavior compared to informal reports of typical behaviors and should not be taken as representative of all cells. This topic deserves much more investigation. Temperature-dependent J–V sweeps were performed from 260–370 K to extract $V_{oc}(T)$. The experimental trends were qualitatively reproduced using a SCAPS-1D model that incorporated literature-based material and defect parameters, with a 10 nm layer at the CdTe/ZnTe interface used to introduce a large enough density of donor-like defects to pin $E_F$. The simulations successfully reproduce the main observables, including the dark/light J–V shape, the degree of dark–light non-superposition, the 1$^{st}$-quadrant JVT roll-over, and the temperature dependence of $V_{oc}$, with only minor adjustments within experimental uncertainty, confirming that an interface-state–induced barrier can reproduce the non-ideal device behaviors.

Figure 1 (a) shows the schematic of the layer structure we implemented in SCAPS-1D; for standard CdSeTe-based cells, the substrate glass would be at the top, and light incident through it. The front side transparent conducting oxide (TCO) is an electron-selective contact of fluorine-doped tin oxide (FTO) of thickness 0.25 µm and degenerate doping. This is followed by a 0.50 µm layer of graded cadmium selenium telluride (CdSeTe) with $E_{gap}$ and other properties linearly graded such that $E_g$=1.4 eV at its front and 1.5 eV at its back. The electron affinity is increased by 0.1 eV at the front of this layer, in accordance with Se, causing $E_g$ to decrease by varying the conduction band edge. We have modeled a p-type doped CdTe layer 1.5-3 µm thick behind the CdSeTe. These layers correspond to standard state-of-the-art deposition order; however, during fabrication, the CdCl$_2$ heat treatment drives recrystallization and Se interdiffusion, distributing Se throughout the absorber but with much lower incorporation in the grain interiors toward the back. Because Se in this region is confined mainly to grain boundaries, the effective bandgap beyond ~500 nm does not decrease as much as it would in a fully graded CdSeTe layer. Thus, representing the back portion as CdTe-like—rather than extending a full linear Se gradient—captures the experimentally observed weak interior Se incorporation and avoids artificially lowering $E_g$. We have additionally simulated a continuously graded CdSeTe layer and it yields nearly identical device behavior, removing doubts about this detail changing any major findings. Last is a 0.25 µm ZnTe hole-selective contact with favorable valence-band alignment to CdTe for hole extraction ($\Delta E_v$=0 to -0.15 eV) but large conduction-band offset to block electron collection [2]. The device parameters used in the simulation are provided in Table I of the Supplementary Material. To understand how the back-interface defects influence hole transport, we first examine the simulated energy-band diagram of CdSeTe/CdTe solar cell shown in Fig. 1 (b). The x-axis spans from an Ohmic contact to ZnTe (0 μm) to the FTO front contact (4 μm). The defective CdTe layer at the ZnTe/CdTe interface has donor-like states slightly below mid-gap but above the p-CdTe $E_F$ in the quasineutral region and has an area density sufficient to produce downward band bending in the adjacent CdTe but not in the $p^+$-ZnTe. This potential barrier for holes acts similarly to (but not exactly like) a parasitic Schottky junction; when the main pn junction is in forward bias, it is in reverse bias and impedes hole extraction. The built-in voltage of this barrier depends on the CdTe doping density and on the defects' number density, energetic distribution, and ratio of hole and electron capture cross sections.

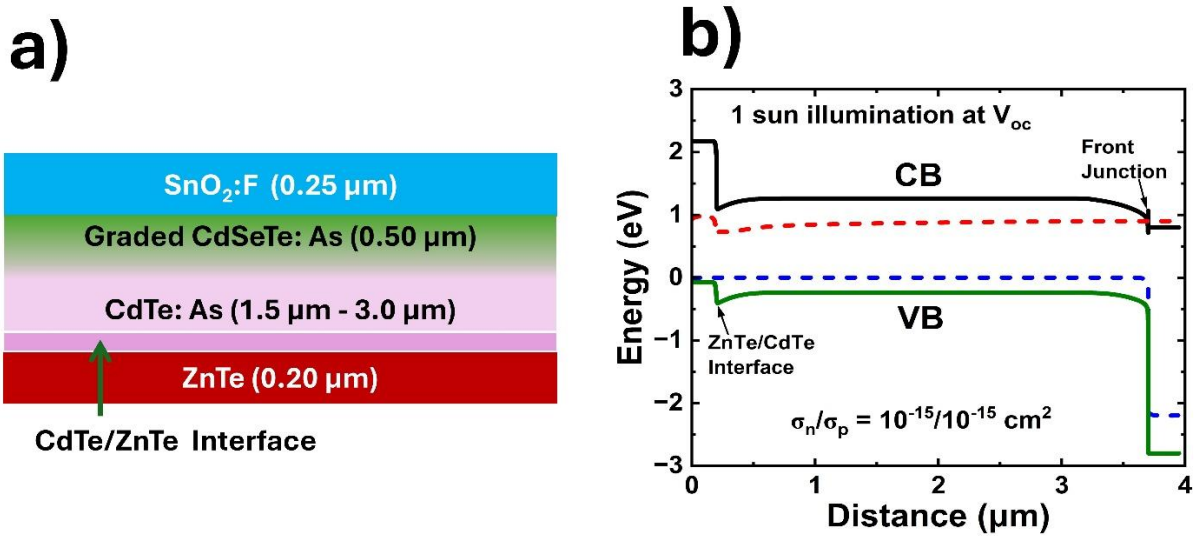

Figure 1: (a) Schematic of the CdSeTe/CdTe thin-film solar cell structure consisting of SnO$_2$:F (0.25 μm), graded CdSeTe:As (0.50 μm), CdTe:As (1.5–3.0 μm), and ZnTe (0.20 μm), highlighting the CdTe/ZnTe back-interface region. (b) Simulated conduction band (CB) and valence band (VB) energy profiles under 1-sun illumination at open-circuit voltage ($V_{oc}$), showing band bending at the ZnTe/CdTe back interface and the front junction.

Like many recent device models [26], we incorporate Urbach tails in optical absorption. However, importantly, each carrier process must obey detailed balance; thus, we explicitly simulate an exponential defect distribution in energy that can trap and thermally detrap carriers (and thus, implicitly, participate in recombination). Other approaches have adopted a smaller bandgap and/or absorbed majority-carrier trapping and detrapping, resulting in an effective mobility lower than the band mobility [27]. Our approach is more physically proper because trapping holes into localized band tail states does not affect the mobility/diffusion constant of the band holes (a property of the extended valence band states themselves) as strongly as it changes the number of valence band carriers. On the other hand, trapping/detrapping of minority electrons protects them from recombination (as is frequently observed as very long apparent time-resolved photoluminescence lifetimes without accompanying high $V_{oc}$) [28] [29], whilst similarly lowering the conduction band conductivity via band electron occupation. The observations of minority electron trapping which produces artificially-long apparent bandgap emission TRPL lifetimes, sub-bandgap radiative recombination centers associated with Se/Te vacancies [4], [30], [31] and/or As doping [32], and majority hole trapping in band tails and or localized states, a high-fidelity model must include both hole and electron traps as well as recombination centers limiting minority carrier lifetime (whether radiative or non-radiative, all recombination at sub-gap energies decreases $V_{oc}$).

There are some subtleties in device modelling of the hole-selective (p-ZnTe) contact: it may be tempting to omit treating this layer explicitly and instead implement a Schottky boundary condition with different recombination velocities for electrons and holes. Experience with state-of-the-art manufactured p-ZnTe layers reveals that their doping may vary from degenerately-doped to as low as $10^{16}$ /cm$^3$ depending on processing conditions. For the lower end of that doping range, series resistance and depletion in the p-ZnTe become significant, and thus it is imperative to include this layer.

The fact that free surfaces of CdTe show photovoltage and the JVT-roll-over and the take-off reported herein both suggest that $E_F$ can be pinned (to a higher or lower degree depending on processing) at the CdTe surface or interface with ZnTe. This can occur with sufficiently high density of donor-like states (since ZnTe and CdTe are p-type) located above the dark $E_F$ in the CdTe (approximately >$10^{12}$ cm$^{-2}$ for doping near $10^{16}$ cm$^{-3}$) [33]. Since the ZnTe/CdTe interface has significant recombination velocity (is not passivated) for both sputtering or thermal evaporation of ZnTe, such defect states likely arise from a combinations of lattice mismatch, sputter-induced damage, and possible chemical contaminants like after CdCl$_2$ processing. Although the physical defect spectrum at such interfaces is expected to extend continuously across the CdTe bandgap, it is well established that this distributed defect population can be represented by an effective amphoteric defect level whose partial occupation stabilizes the Fermi level, resulting in $E_F$ pinning (i.e., $dE_F/d(n,p) \to 0$) [34]. Our model thus contains a 10 nm CdTe defective layer in which we can define additional defect densities of states, e.g a donor-like defect below $E_g/2$ in CdTe with $10^{10} - 10^{13}$ /cm$^2$ equivalent density. Such defect spatial distributions may dynamically change occupation with bias and illumination, thus modulating majority hole extraction. Strictly, an extended spatial distribution of defect states can produce both net charge and a distributed dipole when their occupation varies with distance, whereas a 2D distribution of defects strictly on an interface only provides a delta function of charge in the band diagram. For spatial distributions smaller than the local Debye lengths, this distinction will not impact the overall device physics; however, it may determine the observability of such states in capacitance-based techniques e.g. deep-level transient spectroscopy (DLTS). However, because we wished to investigate any changes in defect occupation vs distance and to use the same model in SCAPS (which does not allow defects at boundary interfaces) for predicting surface photovoltage for samples without the ZnTe layer, we implemented a finite-thickness layer. In our discussion, we define $E_F$ pinning to mean that the dark Fermi level $E_{Fo}$ and the majority (hole) quasi-Fermi level $E_{Fp}$ do not vary appreciably at the interface over the range of voltage and illumination considered: $dE_F/dp \approx 0$. We found that a single donor-like +/0 defect level was sufficient to cause this, and that acceptor-like 0/− defects did not cause pinning. No significant changes in device physics are expected for models incorporating amphoteric +/0/− interface defects or energetic distributions.

**Phenomenology of JV Non-Superposition**

Figure 2(a) compares the dark and illuminated per-cell J–V characteristics obtained from full-string measurements of the same CdSeTe/CdTe mini-module, performed independently at First Solar (solid black) and at the University of Utah (dotted red) under AM 1.5G, 1000 W m$^{-2}$ illumination. The J–V parameters measured after shipping and storage in a N$_2$-purged desiccator at Utah show slight reductions in $V_{oc}$(863 vs. 855 mV) and $J_{sc}$(29.5 vs. 28.3 mA cm$^{-2}$). These small deviations are attributed to the combination of storage and shipping time between the two measurements, and possibly to the slightly hotter and longer wake-up procedure used at First Solar (90 °C, 6 hours in AM1.5). Figure 1(b) plots the same data as for (a) but as $J_{dark}$ and ($J_{light}$-$J_{sc}$): the fact that the later does not overlay the former is the dark–light J–V non-superposition, or according to R. Farshchi "Illuminated JV take-off". Ideally, the current density of an illuminated solar cell ($J_{light}$) should equal the sum of the dark current ($J_{dark}$) shifted downwards by the collected photocurrent. If the collected photocurrent is independent of bias (and is much larger than the diode's dark reverse saturation current $J_o$) it is a constant $J_{sc}$, which would result in the superposition of $J_{dark}$ and ($J_{light} - J_{sc}$) curves. Because the current path is in series in a solar cell, it is difficult to isolate the origins of non-ideal behaviors. Thus we employed our SCAPS modelling to test hypotheses (e.g. defects at the front or the back interface) and determine whether or not they were consistent with experimental observations (no and yes, respectively).

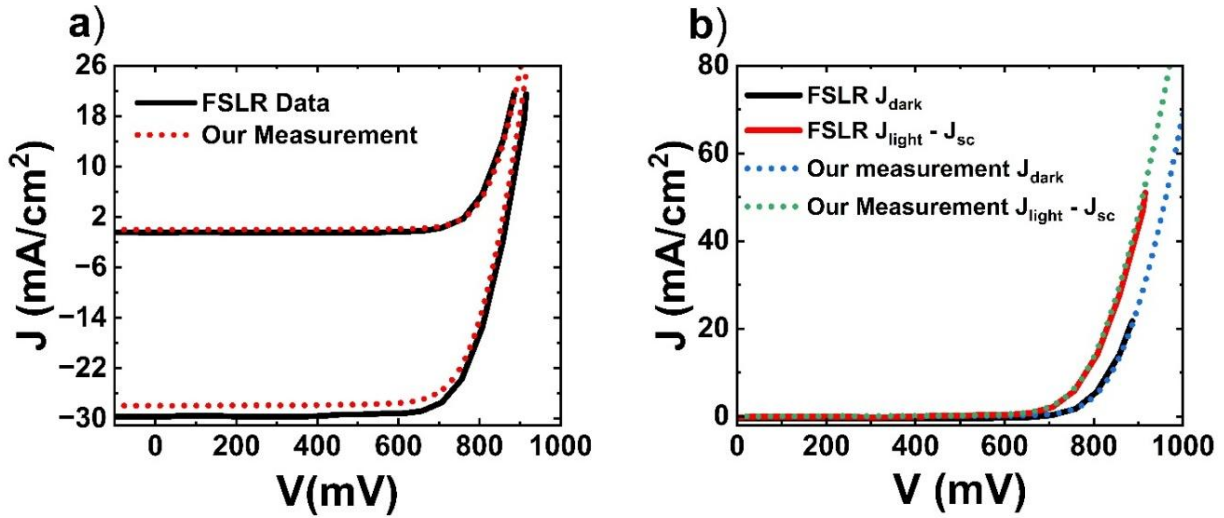

Figure 2: a) Dark and illuminated AM1.5g current–voltage (J–V) characteristics of a mini-module fabricated on production equipment by First Solar. Measurements taken at First Solar are solid black, and those measured at UU are dashed red. Storage time in the unencapsulated state and a different wake-up procedure account for the differences. b) Plot of $J_{dark}$ and ($J_{light} - J_{sc}$) illustrating the JV "take-off" under illumination, which we conclude is caused by a parasitic built-in voltage at the ZnTe/CdSeTe interface that reduces photocurrent collection in forward bias in the dark (as opposed to higher recombination in the light).

**Effects of Defects at the ZnTe/CdTe Interface:**

Figure 3(a) compares the data measured at the University of Utah with a particular SCAPS simulation that matches the dark $J_o$ and ideality factor and $J_{sc}$, FF, and $V_{oc}$ of the light curve. Figure 3(b) presents the corresponding dark J–V curves together and $(J_{light} - J_{sc})$ for the illuminated data, revealing dark–light J–V non-superposition introduced by the built-in voltage at the ZnTe/CdTe interface discussed with Fig. 1. Within the present model, variations in absorber lifetime, acceptor density, and carrier mobility produce a range of simulated device metrics ($V_{oc}$ = 853-896 mV, $J_{sc}$ =23.95-28.3 mA cm$^{-2}$, FF=49.19-76.7%, η=10.1-19.7%) The experimentally measured commercial CdSeTe/CdTe cell (V$_{oc}$=855 mV, J$_{sc}$=28.3 mA cm$^{-2}$, FF=76.4%, η=18.7%) lies within this simulated parameter space, indicating that the model captures the experimentally observed performance trends while explicitly accounting for non-superposition effects.

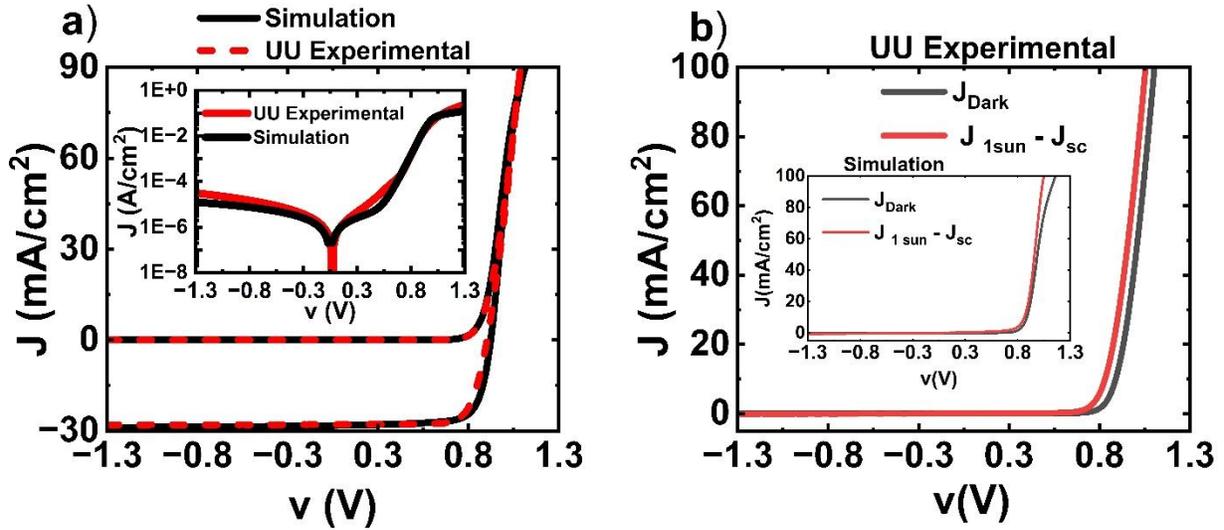

**Figure 3.** Comparison of simulated and University of Utah (UU) experimental J–V characteristics. (a) Dark and 1-sun illuminated J–V curves plotted on a linear current-density scale, showing good agreement between simulation and experimental data. The inset presents the dark J–V characteristics on a semi-logarithmic scale in both forward and reverse bias. (b) Experimental dark current density ($J_{dark}$) and light current density corrected for the short-circuit current ($J_{1sun} - J_{SC}$), highlighting dark–light J–V non-superposition. The inset shows the corresponding simulated $J_{dark}$ and $J_{1sun} - J_{SC}$, reproducing the experimentally observed similar non-superposition behavior.

In Figure 3(a) the SCAPS simulation shown were drawn from within a non-unique parameter space that reproduces the experimentally observed illuminated device behavior and reverse-bias dark current. On a linear scale, variations in absorber lifetime, acceptor density, and carrier mobility yield simulated performance metrics spanning $V_{OC}$ = 853–896 mV, J$_{sc}$ = 23.95–28.3 mA cm$^{-2}$, FF = 49.19–76.7%, and η = 10.1–19.7%. The experimentally measured commercial CdSeTe/CdTe device exhibits V$_{oc}$ = 855 mV, J$_{sc}$ = 28.3 mA cm$^{-2}$, FF = 76.4%, and η = 18.7%, placing it within the simulated parameter space. The inset shows the corresponding dark J–V characteristics on a logarithmic scale, providing a rigorous validation of the simulated dark-current behavior over multiple decades in current density. In the recombination-limited regime, close agreement is observed between experiment and simulation in both magnitude and slope, indicating consistency in the extracted saturation current density and effective ideality factor. Under reverse bias and small forward bias, good but imperfect agreement is achieved by including a finite Ohmic shunt resistance ($R_{sh} = 10^5$ Ω·cm$^2$), which captures the dominant low-voltage leakage pathway. Non-Ohmic shunt behavior (e.g. at grain boundaries or sidewalls of the cells) as well as trapp-assisted tunneling are other possible mechanisms that could not be captured using SCAPS.

Figure 3(b) presents the experimental dark J–V curve together with the quantity ($J_{light} - J_{sc}$) under illumination, revealing pronounced dark–light J–V non-superposition. The inset shows the corresponding simulated curves, demonstrating that the model reproduces the observed non-superposition behavior across the explored parameter space. This behavior originates from the built-in voltage at the ZnTe/CdTe back interface, which modifies carrier transport and collection under illumination relative to the dark condition, as discussed in Fig. 1.

We implemented $E_F$ pinning at the ZnTe/CdTe for trap number densities in the $10^{12}$-$10^{13}$ /cm$^2$ range by determining the defect energy level that yielded zero-bias hole barriers in the range -0.1 to approximately +0.3 eV (approx. $E_{i,CdTe}$-0.4 eV). We make no claim that this is actually the density of states actually on this interface, only that it is sufficient to induce the desired $E_F$ pinning and built-in voltage of the parasitic diode. We carefully investigated whether the occupation of these defects and the associated band bending changed (for a given terminal voltage) with illumination and found a weak dependence of trap occupation and band bending on the ratio of electron-to-hole capture cross sections (the energy level has a much greater effect). This is discussed in detail in the Supplemental Materials. The ratio of cross-sections, however, significantly alters the rear-interface non-radiative recombination behavior, although this is not the dominant source of non-radiative recombination in state of the art devices, in our estimation. The downward band bending actually suppresses interface-trap recombination compared to the flat-band case by starving the process of majority holes. Additionally, the arrival of photogenerated electrons is suppressed for Se alloy disorder-limited electron mobilities (0.1-0.6 cm$^2$/Vs) in line with recent findings from transient grating diffusion experiments [35] and minority carrier lifetimes after the artificial extension, related to trapping/detrapping, is corrected [5]. In thinner cells with higher doping, mobility, and lifetime, the details of recombination at this interface may play a larger role as the detailed-balance limit is approached [10].

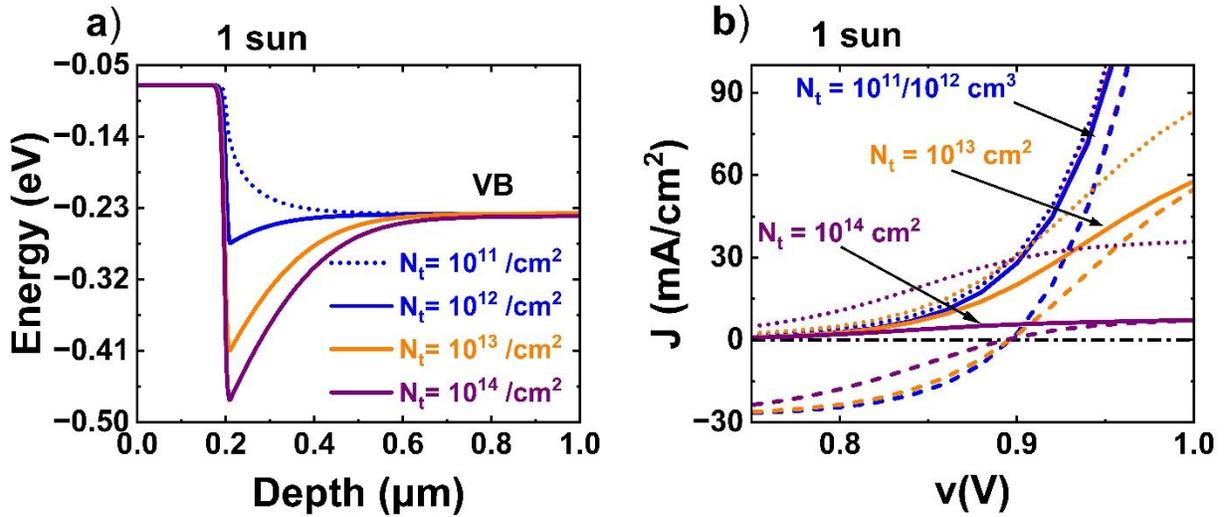

Figure 4. a) Simulated valence band diagram Valence band edge for the variable defect density to show how downward band bending changes with changing defect density in the damage layer. b) The corresponding JV and $J_{light}$-$J_{dark}$ curve represents how JV characteristics change with a changing back interface barrier. In the simulation, ZnTe doping used here is 1E18 cm$^{-3}$.

In Fig. 4, we illustrate how varying the density of interface defects changes the back interface band diagram at $V_{oc}$, as well as the associated dependence of JV characteristics including the emergence of dark-light non-superposition. As expected, Fig. 4a) shows upwards to flat band bending for small interface charge densities and a rather abrupt transition to $E_F$ pinning and downwards bending as density increases [33]. The numerical threshold is roughly near $10^{12}$ /cm$^2$ in our baseline case assuming p in the $10^{15}$ /cm$^3$ range in the CdTe. This is accompanied in Fig. 4b) by the dark and illuminated JV curves' shapes being indistinguishable for low interface defect density, with an onset of progressively more reduced current in forward bias and larger degree of dark-light non-superposition for higher densities. At very high trap density, the phenomenon of 1st quadrant roll-over and near-infinite differential resistance emerges, as would be expected for a parasitic reverse-biased junction [25], [36], [37].

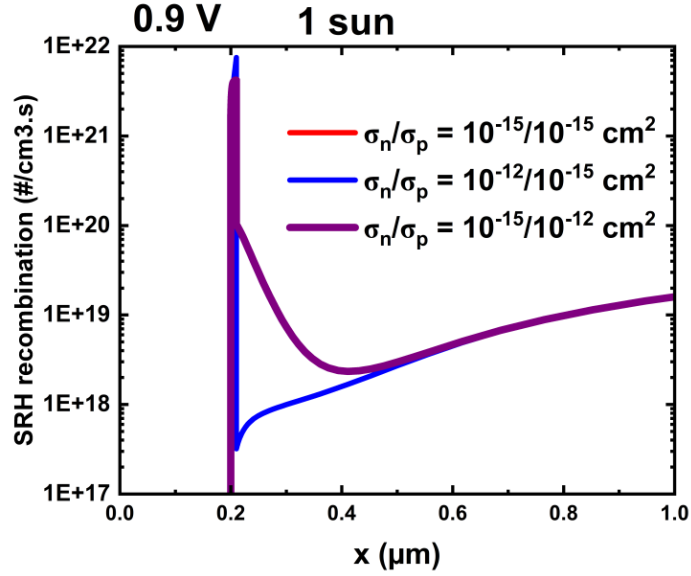

**Figure 5.** Simulated depth-resolved Shockley–Read–Hall (SRH) recombination rate under 1-sun illumination at an applied bias of 0.9 V for different electron-to-hole capture cross-section ratios ($\sigma_n/\sigma_p$) in the back-interface damage layer of the ZnTe/CdTe device. The SRH recombination is strongly localized at the ZnTe/CdTe interface (~0.2 µm) and dominates the recombination throughout the back region of the device, with radiative recombination being much smaller in magnitude (not shown). The curves for $\sigma_n/\sigma_p = 10^{-15}/10^{-15}$ (red) and $10^{-15}/10^{-12}$ (purple) overlap almost exactly over the plotted range, indicating nearly identical SRH recombination profiles.

The behavior of recombination at the interface defects and in the adjacent region of band bending is somewhat counterintuitive. First, we note that the recombination at this interface within our current model comprises approximately 1% of the total recombination in the device for values of the carrier capture cross sections near $10^{-15}$ cm$^2$. The trap-assisted recombination behavior of both interface and bulk defects scales with the product of cross-sections and number densities, so there is no way to distinguish either quantity's value from metrics like dark current and V$_{oc}$ alone. The fact that E$_F$ pinning shows a rather abrupt transition with number density for a given absorber doping does, however, mean that the interface defect density must exceed ~$10^{12}$ /cm$^2$ if JV non-superposition and or 1$^{st}$-quadrant rollover are observed. Additionally, densities above $10^{13}$ /cm$^3$ are physically unreasonable. Once the bulk absorber lifetime is constrained from TRPL experiments, the cross-section of the carrier type limiting the non-radiative recombination rate at the interface can be estimated if the surface recombination velocity can be determined. Intuition from pn junctions and quasineutral regions would suggest that the majority holes should be in excess, and interface recombination limited by minority photoelectrons. However, in this case, the downwards band bending of magnitude $\phi_{back}$ suppresses the hole density and enhances the electron density at the interface by a factor $exp\left(-q\phi_{back}/k_BT\right)$. Shockley-Reed-Hall recombination is maximized when the capture rates of both carrier types are balanced, thus recombination at the CdTe/ZnTe interface is maximized in our model when the ratio $\sigma_n/\sigma_p$ is much greater than 1 (e.g. the blue lines in Fig. 5). When the recombination is maximized at this interface, it lowers the recombination in the adjacent depletion width by reducing the electron concentration – in other words thus recombination in the two regions trade off. These results indicate that the back-interface barrier does not significantly increase recombination; instead, it primarily restricts hole transport. Consequently, the barrier manifests as 1st-quadrant roll-over and FF loss through a transport-limited, series-resistance-like mechanism rather than through enhanced recombination.

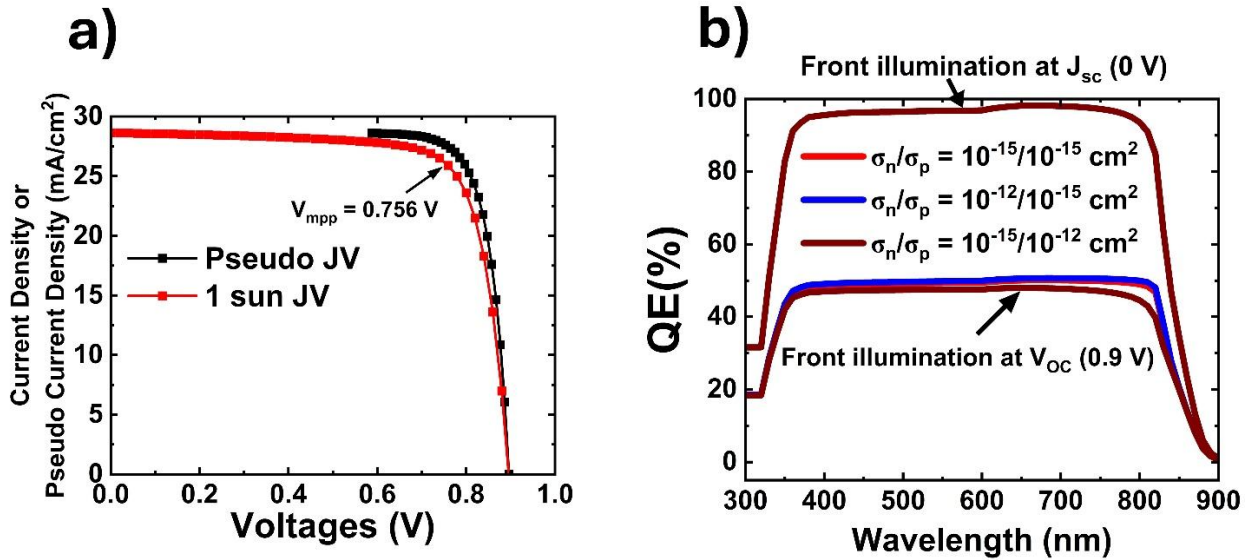

Figure 6. (a) Comparison of the measured 1-sun J–V curve and the corresponding pseudo-J–V curve of the CdSeTe/CdTe device, illustrating voltage-dependent current collection and a maximum power point at $V_{mpp}$ = 0.756V. The pseudo J-V is constructed from measurements taken under reduced illumination intensities (0.68–1 sun). (b) Simulated external quantum efficiency (QE) spectra under front illumination at short-circuit (0 V) and open-circuit (0.9 V) conditions for different electron-to-hole capture cross-section ratios ($\sigma_n/\sigma_p$) at the back-interface damage layer, highlighting the voltage-dependent collection behavior associated with the back-interface barrier.

To further investigate the impact of the effective series resistance associated with the pinned back interface, we employ the pseudo-J–V method derived from illumination-dependent $J_{SC}$–$V_{OC}$ measurements, a widely used approach to isolate intrinsic diode behavior and quantify transport-related losses in thin-film solar cells [38]. The pseudo-current density is defined as $J_{\text{pseudo}} = J_{\text{SC, 1 sun}} - J_{\text{SC}}(\text{suns})$, which removes series-resistance contributions and reveals the ideal diode response. Using this method, the illumination-dependent $J_{\text{SC}}$–$V_{\text{OC}}$ data are used to construct the pseudo-J–V curve and compared with the simulated 1-sun J–V in Fig. 6(a). A measurable separation between the curves appears prior to $V_{\text{OC}}$ due to illumination-dependent recombination and transport effects; however, the dominant deviation occurs near $V_{\text{OC}}$, where the simulated J–V falls below the pseudo-J–V. Because the pseudo-J–V excludes resistive losses, this divergence at high forward bias demonstrates that back-contact-induced series resistance, rather than recombination, is the primary origin of the observed fill-factor degradation.

Building on this pseudo-J–V analysis, Fig. 6(b) examines the corresponding impact of the back-interface barrier on voltage-dependent carrier collection through the internal quantum efficiency (IQE). At short-circuit conditions (0 V), the IQE exceeds 90% for all electron-to-hole capture cross-section ratios ($\sigma_n/\sigma_p$), indicating efficient carrier extraction under strong internal fields and minimal sensitivity to interface asymmetry. Under forward bias near $V_{\text{OC}}$, however, the IQE becomes strongly dependent on $\sigma_n/\sigma_p$. The largest IQE degradation occurs when $\sigma_p > \sigma_n (10^{-15}/10^{-12}$ cm²), a moderate reduction appears in the symmetric case ($10^{-15}/10^{-15}$ cm²), and the smallest degradation occurs when $\sigma_n > \sigma_p (10^{-12}/10^{-15}$ cm²). This ordering is consistent with the interfacial charge and barrier profiles summarized in Table II of supplementary information: although the symmetric and $\sigma_n > \sigma_p$ cases exhibit similar band bending, the latter has a slightly lower integrated positive charge, resulting in a weaker hole barrier and higher IQE. Notably, the $\sigma_n/\sigma_p = 10^{-12}/10^{-15}$ cm² case shows the highest recombination in the damage layer yet the highest IQE near $V_{\text{OC}}$, confirming that IQE loss is governed not by SRH recombination magnitude but by the strength of the back-interface hole barrier under reduced internal fields. Together, Fig. 6 (a) and 6 (b) demonstrate that the pinned back-interface barrier, rather than recombination alone, dominates the voltage-dependent collection behavior and the associated fill-factor loss.

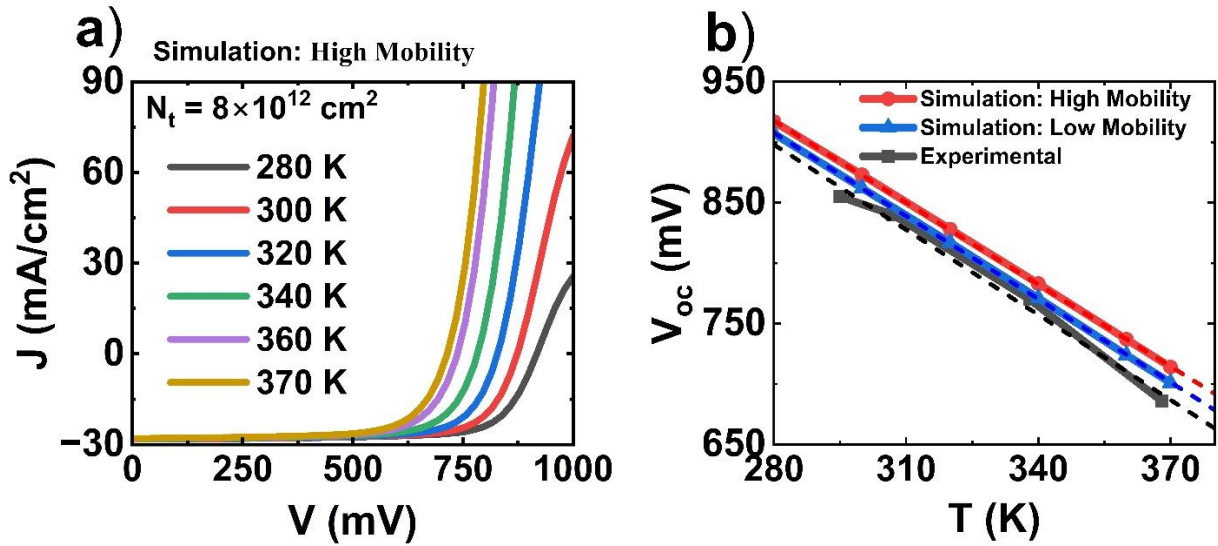

Figure 7: Temperature-dependent current density–voltage (J–V) characteristics and extracted open-circuit voltage ($V_{OC}$) for a state-of-the-art CdSeTe/CdTe solar cell. (a) Simulated J–V curves for the high-mobility ($\mu_n/\mu_p$ = 100/15 cm$^2$/V·s) case over 280–370 K, showing first-quadrant roll-over associated with the back-interface barrier at lower temperatures (≤ 300 K). (b) $V_{OC}$ as a function of temperature, comparing experimental data (gray) with simulations assuming high mobility ($\mu_n/\mu_p$ = 100/15 cm$^2$/V·s, red) and low mobility ($\mu_n/\mu_p$ = 2/0.5 cm$^2$/V·s, blue). Linear fits yield activation energies of 1.55 eV (experiment) and 1.54 eV (simulation), consistent with the CdSeTe bandgap and indicating thermally activated recombination as the dominant $V_{OC}$-limiting mechanism.

Next, we examine the temperature dependent JV and corresponding $V_{OC}$ behavior. Fig. 7(a) shows the JVT data and Fig. 7 (b) compares the temperature-dependent $V_{OC}$ of a state-of-the-art CdSeTe/CdTe device with SCAPS simulations. Both experimental and simulated curves show an approximately linear decrease of $V_{OC}$ with temperature, consistent with thermally activated band-edge recombination as the dominant voltage-limiting mechanism. The extracted temperature coefficients are similar (≈ −2.6 mV K$^{-1}$), and linear extrapolation yields activation energies of 1.55 eV (experiment) and 1.54 eV (simulation), in closely match with the CdSeTe absorber bandgap. The simulated J–V curves further reveal pronounced first-quadrant roll-over in the high-mobility ($\mu_n/\mu_p$ = 100/15 cm$^2$/V·s ) case at lower temperatures (≤ 300 K), indicating the presence of a back-interface potential barrier. The requirement of a relatively high defect density to induce strong roll-over in this high-mobility regime highlights that the effect is not transport-limited in the bulk, but instead arises from a pinned back interface that maintains band bending under forward bias. Together, the $V_{OC}$–T behavior and the temperature-dependent roll-over consistently indicate that back-interface-induced band bending—rather than bulk recombination alone—governs the temperature-dependent device response and forward-bias transport limitations.

**Conclusions:** This work demonstrates that the characteristic dark–light JV non-superposition, voltage-dependent long-wavelength QE loss, and JVT roll-over observed in CdSeTe/CdTe solar cells originate from a back-interface potential barrier created by Fermi-level pinning at the CdTe/ZnTe contact. A SCAPS model calibrated to state-of-the-art device behavior reproduces these signatures through a donor-like defect layer at the back interface, which induces strong, bias-dependent downward band bending and governs forward-bias carrier transport. Temperature-dependent $V_{oc}(T)$ measurements yield activation energies close to the absorber bandgap, indicating that bulk or band-edge recombination determines the open-circuit voltage, whereas the pinned back interface primarily affects forward-bias transport and fill-factor loss. Pseudo-JV analysis further confirms that the series resistance associated with the pinned interface dominates the FF degradation. Simulations with a controlled Schottky barrier show that any sufficiently strong back-interface barrier can produce similar JV non-superposition and take-off behavior; however, the combined evidence—including voltage-dependent QE, JVT roll-over, interfacial charge behavior, and mobility–defect trends—points to defect-mediated Fermi-level pinning as the physically consistent mechanism in CdSeTe/CdTe devices. Finally, our simulations suggest that an ultrathin DLC layer may modestly reduce interfacial defects and weaken Fermi level pinning.

**Acknowledgments:** This work was based [in part] on research sponsored by the U.S. Department of Energy Office of Energy Efficiency and Renewable Energy Solar Energy Technologies Office agreement number 37989 through National Laboratory of the Rockies, operated under Contract No. DE-AC36-08GO28308. The views expressed in the article do not necessarily represent the views of the DOE or the U.S. Government. The U.S. Government retains and the publisher, by accepting the article for publication, acknowledges that the U.S. Government retains a nonexclusive, paid-up, irrevocable, worldwide license to publish or reproduce the published form of this work, or allow others to do so, for U.S. Government purposes.